\documentclass{aa}
\usepackage[varg]{txfonts}
\usepackage{graphicx}
\usepackage{txfonts}
\usepackage{natbib}
\bibpunct{(}{)}{;}{a}{}{,} 
\begin{document}
\title{On the existence of warm H-rich pulsating white dwarfs}

\author{Leandro G. Althaus\inst{1,2}, 
        Alejandro H. C\'orsico\inst{1,2}, 
        Murat Uzundag\inst{3},   
        Maja Vu\v{c}kovi\'{c}\inst{3},
        Andrzej S. Baran\inst{4},
        Keaton J. Bell\inst{5,6},
        Mar\'ia E. Camisassa\inst{1,2},
        Leila M. Calcaferro\inst{1,2},
        Francisco C. De Ger\'onimo\inst{1,2}, 
        S. O. Kepler\inst{7}, 
        Roberto Silvotti\inst{8}
        }
\institute{Grupo de Evoluci\'on Estelar y Pulsaciones. 
           Facultad de Ciencias Astron\'omicas y Geof\'{\i}sicas, 
           Universidad Nacional de La Plata, 
           Paseo del Bosque s/n, 1900 
           La Plata, 
           Argentina
           \and
           IALP - CONICET
           \and
           Instituto de F\'isica y Astronom\'ia, Universidad de Valparaiso, Gran Breta\~na 1111, Playa Ancha, Valpara\'iso 2360102, Chile
           \and
           Uniwersytet Pedagogiczny, Obserwatorium na Suhorze, ul. Podchor\c{a}\.zych 2, 30-084 Krak\'ow, Polska
           \and 
           DIRAC Institute, Department of Astronomy, University of Washington, Seattle, WA 98195-1580, USA
           \and
           NSF Astronomy and Astrophysics Fellow and DIRAC Fellow
           \and               
           Instituto de F\'{i}sica, Universidade Federal do Rio Grande do Sul, 91501-900, Porto-Alegre, RS, Brazil
           \and
           INAF – Osservatorio Astrofisico di Torino, strada dell’Osservatorio 20, 10025 Pino Torinese, Italy
           }
\date{Received ; accepted }

\abstract{The possible existence of warm ($T_{\rm eff}\sim19\,000$ K) pulsating DA white dwarf (WD) stars, hotter than ZZ Ceti stars,
 was predicted in theoretical studies more than 30 yr ago. 
 These studies reported the occurrence of  $g$-mode pulsational 
instabilities due to the $\kappa$ mechanism acting in the partial ionization zone of He below the H envelope in models of DA WDs with very thin H envelopes ($M_{\rm H}/M_{\star} \lesssim 10^{-10}$). However, to date, no pulsating warm DA WD has been discovered, despite the  varied theoretical and observational evidence suggesting that a fraction of WDs should be formed with a range of very low H content. }
{We  re-examine the pulsational predictions for such WDs on the basis of new full evolutionary sequences. We analyze all the warm DAs observed by TESS satellite up to Sector 9 in order to search for the possible pulsational signal. }
{We compute  WD evolutionary  sequences of masses 0.58 and 0.80 $M_\sun$ with H content in the range 
$-14.5 \lesssim \log(M_{\rm H}/M_{\star}) \lesssim -10$,  appropriate  for  the   study  of pulsational instability  of warm DA WDs. Initial models were extracted
from progenitors that were evolved through  very late thermal pulses on the early cooling branch. We use {\tt LPCODE} stellar code to which we have incorporated a new  full-implicit  treatment  of time-dependent  element  diffusion for precisely modeling  the H/He transition zone in evolving  WD models with very low H content.  The non-adiabatic pulsations of our warm DA WD models were computed in the effective  temperature range  of $30\,000-10\,000$ K, focusing  on $\ell= 1$ $g$ modes  with periods  in the range $50-1500$ s.}
{We  find  that  traces of H surviving the very late thermal pulse float to the  surface, eventually forming growing, thin pure H envelopes and rather extended H/He transition zones. We find that such extended transition zones inhibit the excitation of $g$ modes due to partial ionization of He below the H envelope. Only in the case that the H/He transition is assumed much more abrupt than predicted by  diffusion, models do exhibit pulsational instability. In this case, instabilities are found only in WD models with H envelopes in the range of $-14.5 \lesssim \log(M_{\rm H}/M_{\star}) \lesssim -10$ and at effective temperatures  higher than those typical of ZZ Ceti stars, in agreement with  previous studies. None of the 36 warm DAs observed so far by  TESS satellite are found to pulsate.}
{Our study suggests that the non-detection of pulsating warm DAs, if WDs with very thin H envelopes do exist, could be attributed to the presence of a smooth and extended H/He  transition zone. This could be considered as an indirect proof that element diffusion indeed operates in the interior of WDs.}

\titlerunning{Warm H-rich pulsating white dwarfs}
\authorrunning{Althaus et al.}

\maketitle

\section{Introduction}
\label{introduction}

The existence of various types of pulsating white dwarf (WD) stars characterized by multiperiodic luminosity variations due
to $g$(gravity)-mode pulsations has been well established by theoretical and observational evidence \citep{2008ARA&A..46..157W,2008PASP..120.1043F,2010A&ARv..18..471A}. During their
lives, WDs experience at least one stage of pulsational instability, during which  peak-to-peak
amplitudes  between 0.1 mmag and 0.4 mag in optical light
curves are expected. Among the different types of observed pulsating WDs \citep{2019A&ARv..27....7C} we mention the most abundant ones: 
{\it (i)}  the hydrogen (H)-rich variables ZZ Ceti
or DAVs \citep{1968ApJ...153..151L} at low effective temperatures and high gravities  ($10\,400$ K $\lesssim
T_{\rm eff} \lesssim  12\,400$ K and $7.5 \lesssim \log g \lesssim
9.1$); {\it (ii)} the variables V777 Her  or DBVs, with almost pure helium (He) atmospheres, spanning a range in effective
temperature and surface gravity of $22\,400$ K $\lesssim T_{\rm eff} \lesssim 32\,000$ K and $7.5
\lesssim \log  g \lesssim  8.3$,  theoretically  predicted  by  \cite{1982ApJ...252L..65W} before their
discovery \citep{1982ApJ...262L..11W}; {\it (iii)}
the ``hot DAVs'' \citep[$T_{\rm eff} \sim 30\,000$ K, $7.3
  \lesssim \log  g \lesssim 7.8$;][]{2008MNRAS.389.1771K,2013MNRAS.432.1632K}, 
  whose existence was anticipated by the theoretical calculations of \cite{2005EAS....17..143S,2007AIPC..948...35S}; {\it (iv)} the pulsating PG1159 stars or GW Vir variable
stars, after the prototype of the class, PG 1159$-$035
\citep{1979wdvd.coll..377M}, rich in He, carbon (C), and oxygen (O), being 
the hottest known
class of pulsating WDs and pre-WDs ($80\,000$ K $\lesssim T_{\rm eff}
\lesssim 180\,000$ K  and $5.5 \lesssim \log g \lesssim  7.5$); and {\it (v)}
the H-rich ELMVs (Extremely Low-Mass WDs variable) with  $7\,800$ K $\lesssim
T_{\rm eff} \lesssim 10\,000$ K  and $6 \lesssim \log g \lesssim
6.8$ \citep{2012ApJ...750L..28H} and {\it (vi)} the He/H-atmosphere pre-ELMVs ($8\,000$ K $\lesssim  T_{\rm eff} \lesssim 13\,000$ K  and $4 \lesssim \log g \lesssim  5$), the probable precursors of ELMVs \citep{2013Natur.498..463M}.

In addition to these pulsating WDs, early theoretical calculations predicted the existence of pulsating 
DA WDs  hotter than ZZ Ceti stars, hereinafter warm DAVs, expected to be found at effective temperatures 
comparable to those of the cool edge of the  DBV instability strip. 
In fact, theoretical investigations carried out by \cite{1982ApJ...252L..65W} 
\citep[see, also,][]{1982PhDT........27W} led to the discovery of $g$-mode pulsational 
instability due to $\kappa-$mechanism operating in the  partial ionization zone of He below the H envelope
in models of DA WDs harboring 
very thin H envelopes ($M_{\rm H}/M_{\star} \lesssim 10^{-10}$), for effective 
temperatures of $\sim 19\,000$~K. A major issue of this finding was related to the possibility of
constraining the mass of the residual H envelope left in a WD by eventually  observing pulsating WDs near this temperature  \citep{1982ApJ...252L..65W}.

The existence of WDs with such very thin H envelopes is not discarded neither by the theory of stellar evolution nor
by observations. From the observational point of view, nearly 80\% of the spectroscopically identified WDs
are characterized by H-rich atmospheres. Systematic spectroscopic and asteroseismological studies
indicate that between 15\% and 20\% of such WDs are expected to harbor  thin
H envelopes with $M_{\rm H} \lesssim 10^{-6} M_\sun$ 
\citep[see][]{2008ApJ...672.1144T,2009MNRAS.396.1709C,2012MNRAS.420.1462R}. On the other hand, single stellar evolution
theory predicts that usually WDs should be formed with a total H content of about $M_{\rm H} \sim 10^{-3} -10^{-5}M_\sun$, depending on the stellar mass and metallicity of the progenitor 
\citep{2015A&A...576A...9A}. However, certain evolutionary scenarios predict
the formation of DA WDs with thin H envelopes. One of such scenario involves the occurrence of very late thermal pulses (VLTP) that take place  when
the WD progenitor experiences its  last thermal pulse on the
early WD cooling branch. In this case, the H-rich envelope is
ingested by the He-shell flash convective zone, where H is
burnt in the hot interior of the star
\citep{1995LNP...443...48I,1999A&A...349L...5H}. Recent full evolutionary calculations of the event by
\cite{2017ASPC..509..435M} predict the formation of DA WDs with H content of $M_{\rm H}\lesssim 10^{-7} M_\sun$. 
In line with this, 
if the last  thermal pulse happens during the horizontal evolution of the post-AGB star in
the HR diagram, a scenario termed Late Thermal Pulse (LTP) occurs \citep{2001Ap&SS.275....1B}.  Here, the H-rich envelope is not
burned but diluted by the deepening of the convective  envelope once
the star evolves back to the AGB after the LTP. As shown  by
\citet{2005A&A...440L...1A}, the H diluted into the deeper parts of the
envelope  is later burned as the H-deficient  star contracts
again to the WD cooling track,  leading to the formation of WDs with a
low H-content ($M_{\rm H} \lesssim 10^{-6}-10^{-7} M_\sun$).
For the case of VLTP remnants of masses  $M_{\rm WD}\gtrsim 0.6 M_\sun$,
\cite{2017ASPC..509..435M} obtained DA WDs with H content of the order of $10^{-11} M_\sun$. On the other hand, the existence of WDs with extremely
thin H envelopes (of the order of $M_{\rm H} \lesssim 10^{-14} M_\sun$) has been invoked in the context of the
spectral evolution that some WDs experience as they evolve, see \cite{1987fbs..conf..319F,2007AIPC..948...35S}.
Finally, additional observational evidence supporting the picture that a fraction of DA WDs could harbour thin H
envelopes is provided by a new dynamical mass determination of WD 40 Eri B, for which a fractional H envelope mass
of  $\approx 10^{-10}$ has recently been inferred by \cite{2017ApJ...848...16B} 
\citep[see, also,][]{2019MNRAS.484.2711R}.
All this evidence suggests that WDs could be formed
with  H envelopes orders of magnitude less massive than the canonical value of  $M_{\rm H} \approx 10^{-4} M_\sun$. 

In view of these considerations, and despite the numerous   discoveries of pulsating WD stars in the last decade, both from the ground and from space, it is surprising that no warm DAV WD has been detected thus far. Ground-based observations, particularly carried out  with the spectral  observations of the Sloan
Digital Sky Survey \citep[SDSS;][]{2000AJ....120.1579Y},  have
increased the number of known WDs by a factor of $15$
\citep{2013ApJS..204....5K,2016MNRAS.455.3413K,Kepler18,2017EPJWC.15201011K,
2019MNRAS.482.4570G} and the number 
of pulsators by a factor of $4$ \citep{2004ApJ...607..982M,2005ApJ...625..966M,2006A&A...450..227C,2007ASPC..372..583V,
  2009ApJ...690..560N,2013MNRAS.430...50C}. On the other hand, the {\it  Kepler} satellite observations, both main mission
\citep{2010Sci...327..977B} and K2 \citep[two-wheel   operation,][]{2014PASP..126..398H}, increased the number of known
WD pulsators by a factor of 2 \citep{2017ApJS..232...23H}. The fact that no warm DAV has emerged from the various surveys 
has prompted us to re-examine the pulsational properties of warm DAV WDs on the basis of detailed 
WD evolutionary sequences that incorporate a full implicit treatment  of time-dependent element diffusion, appropriate for  modeling the formation of the H-rich envelope and H/He transition  in WDs with very thin H envelopes resulting from post-VLTP remnants. In this study, we also examine the light curves provided by TESS satellite (Ricker et al., 2015) to search for possible candidates of warm DAV WDs.

The paper  is organized  as follows.  In Sect.~\ref{tess} we 
present null results in the search for warm DAVs on the basis of a sample 
of DA WDs with effective temperatures between $\sim 17\,500$ K and 
$\sim 22\,500$ K observed with TESS. In Sect.~\ref{code}  we  
briefly describe some details of our stellar and pulsational codes, and evolutionary sequences, and present in detail our  treatment of 
time-dependent element diffusion. In Sect.~\ref{results}  we present  
the predictions of our evolution and pulsation analyses.  Finally,  in  Sect.~\ref{conclusions}  we  summarize  the  main
findings of the paper, and we elaborate on our conclusions.

\section{Observational insights from TESS}
\label{tess}

We have searched for pulsations in warm DA WDs observed by the recent NASA Space Mission of Transiting Exoplanet Survey Satellite \citep[TESS;][]{Ricker2014}. To this end, we have cross-checked the targets from the Montreal White Dwarf Database (MWDD)\footnote{http://www.montrealwhitedwarfdatabase.org} \citep{Dufour2017}, which provides information about all known WDs, with the targets observed with TESS. We found 36 objects matching our criteria of $T_{\rm eff}$ between $17\,500$ and $22\,500$ K. They are listed in Table \ref{table1} of the Appendix \ref{appendixA}\footnote{We note that stars fainter than $T_{\rm mag}= 17$ have detection thresholds not useful for detecting pulsations.}.

We used the short-cadence (SC) observations sampled every 2 minutes, which allows us to analyze the frequency range up to the Nyquist frequency at around ~4167 $\mu$Hz. 
We downloaded all available data (sector 1 to 13) from the “Barbara A. Mikulski Archive for Space Telescopes” (MAST)\footnote{http://archive.stsci.edu}. The data are in the FITS format which includes all the photometric information, which have been already processed with the Pre-Search Data Conditioning Pipeline \citep{2016SPIE.9913E..3EJ} to remove common instrumental trends. From the FITS file, first we have extracted times in barycentric corrected Julian days (“BJD - 245700”), and fluxes (“PDCSAP FLUX”). In order to increase the signal-to-noise ratio (S/N) of the data in the Fourier space, we detrended the light curves after $\sigma$ clipping, removing the outliers that vary significantly from the local standard deviation ($\sigma$) by applying a running $4\sigma$ clipping mask. The fluxes were then normalized and transformed to amplitudes in parts-per-thousand (ppt) unit (($\Delta I$/$I-1)\times 1000$). 

In order to search for periodicity, we assessed the Fourier Spectrum of each light curve. Then, we calculated the median noise level of each data set ($\sigma$) and used $5\sigma$ as the detection threshold \citep{Baran2015}. The 
amplitude spectra of all DA WDs analyzed are presented in Figs. \ref{ft_1} to \ref{ft_6}, where the blue line represents the $5\sigma$ threshold. As can be seen, none of the 36 DA WDs depicted in Figs. \ref{ft_1} to \ref{ft_6} have any signal above the threshold in the frequency region of interest. Therefore, we conclude that this sample of DA WDs observed with TESS (in sectors 1 to 9) does not show any signs of pulsations up to the Nyquist frequency at ~4200 $\mu$Hz.   

\section{Evolutionary/pulsational models and numerical inputs}
\label{code}

Our pulsation calculations are based on full
WD evolutionary models of different  stellar masses. Initial models were 
extracted from  progenitor star models that were evolved from the ZAMS, to the thermally-pulsing AGB phase, and through the VLTP occurring at the beginning of the cooling branch, \cite{2006A&A...454..845M}.  VLTP is one of the most plausible
single-evolution scenarios that predicts the formation of hydrogen-poor WDs \citep[see][for recent discussion]{2019A&ARv..27....7C}. To encompass the range of H envelope mass that should be expected in
warm DAV WDs, the H content of our post VLTP sequence has been artificially reduced, mimicking a more efficient hydrogen mixing and burning during the VLTP.  Specifically, we concentrate on warm DAV WD models of masses $0.58$ and $0.80  M_\sun$ with H content
in the range  $-14.5 \lesssim \log(M_{\rm H}/M_{\star}) \lesssim -10$.

We use our stellar evolutionary code,  {\tt    LPCODE}, to follow the evolution of our selected
post-VLTP remnants
from the very early stages of WD evolution down to an effective temperature of 10,000K, well below
the expected domain of warm DAV WDs.  {\tt    LPCODE} is a well tested and calibrated stellar code;
see  
\cite{2003A&A...404..593A},                \cite{2005A&A...435..631A},
\cite{2015A&A...576A...9A},      and
\cite{2016A&A...588A..25M}  for details.  {\tt    LPCODE}
has been widely used to study the evolution of low-mass and WD  stars  ---    see    \cite{2008A&A...491..253M},
\cite{2010Natur.465..194G},                \cite{2010ApJ...717..897A},
\cite{2010ApJ...717..183R},                \cite{2011ApJ...743L..33M},
\cite{2011A&A...533A.139W},       \cite{2012MNRAS.424.2792C},
\cite{2013A&A...557A..19A}, and
\cite{2016A&A...588A..25M}.

We compute the non-adiabatic pulsations of our warm DA WD models in the effective-temperature range
$30,000\ {\rm K}-10,000$ K, amply covering the region where warm DAV WDs should be expected. We have focused on $\ell= 1$ $g$ modes with periods in the range $50-1500$ s. To this end, we employ the non-adiabatic version of the {\tt LP-PUL} pulsation code, which is described in  \cite{2006A&A...458..259C}. 

\subsection{Numerical treatment of time-dependent element diffusion}

After the violent proton  burning  and mixing during the VLTP episode, only tiny vestiges of H remain, extending from the surface down to as deep as $3 \times 10^{-4} M_\sun$. As WD evolves along the cooling branch, such traces of H will float to the surface as a result of gravitational settling, thus  forming an increasing pure (and thin) H envelope. 
We have explored two situations: a discontinuous H/He transition separating the pure H envelope from the underlying pure He buffer, and a H/He interface as predicted by  element diffusion. We have implemented a new treatment for time-dependent element diffusion, details of which can be found in Appendix \ref{appendixB}.

\begin{figure}
	\centering
	\includegraphics[width=1.0\columnwidth]{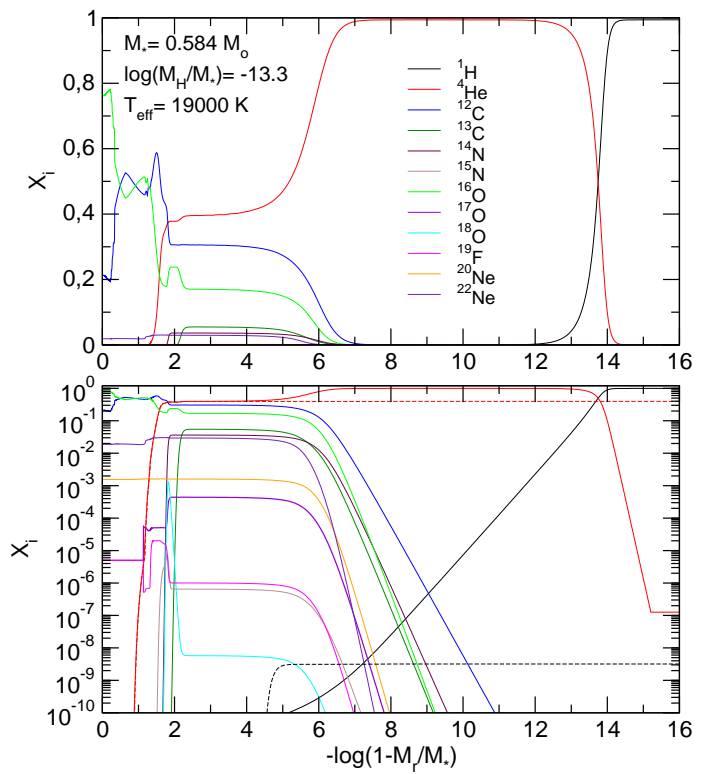}
	\caption{Upper panel: fractional abundance by mass of selected nuclear species in terms of the outer mass fraction, corresponding to a DA WD model characterized by $M_{\star}=0.584 M_{\sun}$, $T_{\rm eff}= 19\,000$ K and $\log(M_{\rm H}/M_{\star})= -13.3$. Lower panel: same as in the upper panel, but with a logarithmic scale for the chemical abundances. In this panel, the black and red dashed lines correspond to initial fractional abundances of H and He, respectively.} 
	\label{xi-2panels}
\end{figure}

As an illustrative example of the performance of our new treatment of element diffusion, in Fig. \ref{xi-2panels} we show the  abundances by mass ($X_i$) of selected chemical species in terms of the external fractional mass, for a DA WD model  
with $M_{\star}= 0.584 M_{\sun}$, $T_{\rm eff}= 19\,000$ K and $\log(M_{\rm H}/M_{\star})= -13.3$. The upper panel shows the abundances on a linear scale, thus emphasizing the dominant species, while the lower panel shows the abundances on a logarithmic scale,  making it possible to clearly visualize the less abundant species. In the lower panel, the black and red dashed lines are associated to the initial profiles of H and He, respectively. Element diffusion predicts a wide and smooth H/He transition zone. Note the tail of H digging into deeper layers as a result of
element diffusion. The imprints on the chemical abundance distribution left by the violent H burning during VLTP are visible, in particular the high abundance of $^{13}$C. Note also the presence of a thick intershell rich in He and carbon, left by the convection zone driven by the last He thermal pulses.

\section{Evolutionary and pulsational results}
\label{results}

\begin{figure}
	\centering
	\includegraphics[width=1.0\columnwidth]{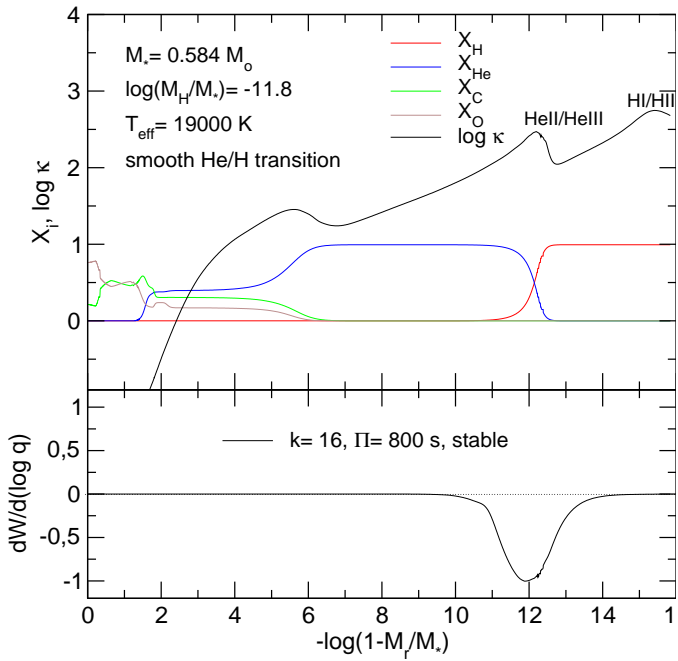}
	\caption{Upper panel: inner chemical  profiles  of  $^{16}$O, $^{12}$C, $^{4}$He  and H, and 
	the logarithm of the Rosseland opacity ($\kappa$) versus the outer mass fraction coordinate,
	corresponding to a DA WD model with $M_{\star}= 0.584 M_{\sun}$, $\log(M_{\rm H}/M_{\star})= -11.8$, and $T_{\rm eff}= 19\,000$ K.
        The H/He interface is shaped by element diffusion. The bumps in $\kappa$ due to the second ionization of He and the
        ionization of H are evident. Lower panel: the normalized differential work function corresponding to a stable $g$ mode
        with $k= 16$ and period $\Pi= 800$ s.} 
	\label{058-12-19000-difu}
\end{figure}

\begin{figure}
	\centering
	\includegraphics[width=1.0\columnwidth]{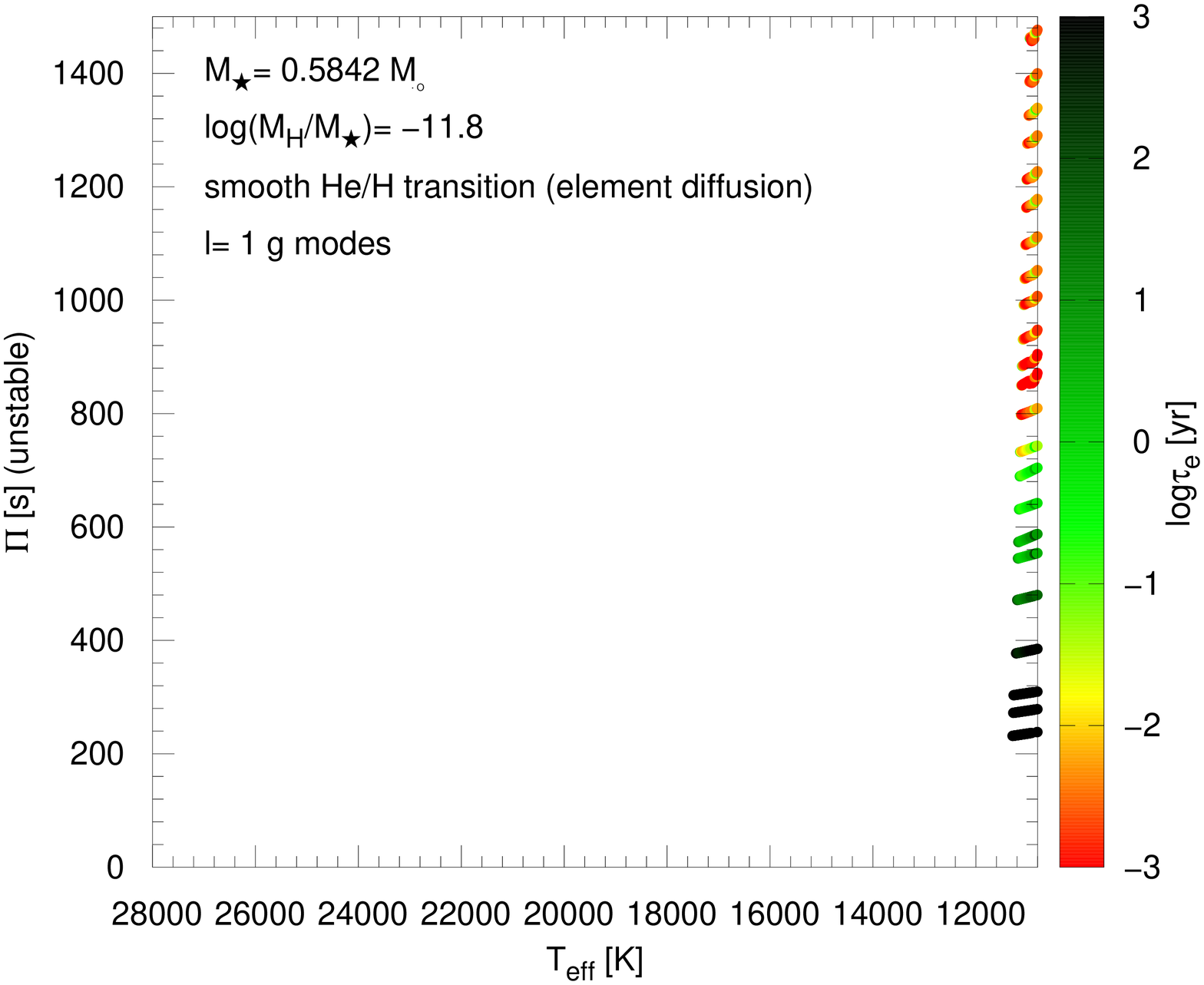}
	\caption{Unstable $\ell=1$ $g$-mode periods $(\Pi)$ in terms of the effective temperature for the sequence of DA WD models with $M_{\star}= 0.584 M_{\sun}$ and $\log(M_{\rm H}/M_{\star})= -11.8$.  Color coding indicates the value of the logarithm of the $e$-folding 
	time ($\tau_{\rm e}$) of each unstable mode (right scale). Only unstable modes for effective temperatures characteristic of the ZZ Ceti stage ($T_{\rm eff} \lesssim 11\,500$ K) are found.} 
	\label{per-teff-eft-059-12}
\end{figure}

In the upper panel of Fig. \ref{058-12-19000-difu} we depict the chemical profiles of $^{16}$O, $^{12}$C, $^{4}$He, H, and the Rosseland opacity as a function of the
coordinate $-\log(1-M_r/M_{\star})$ for a template model with  $M_{\star}= 0.584 M_{\sun}$, $\log(M_{\rm H}/M_{\star})= -11.8$, and 
$T_{\rm eff}= 19\,000$ K. As in the situation analysed in Fig. \ref{xi-2panels}, we also find  that element diffusion leads to
an extended  H/He transition region, with  
the tail of the H profile reaching layers well below the point at which $X_{\rm He}= X_{\rm H}= 0.5$. 
Note the presence of bumps in the Rosseland opacity 
corresponding to the second ionization of He ($\log T\sim 5.3$) and the ionization of 
H ($\log T \sim 4.58$). In particular, the HeII/HeIII bump is quite pronounced 
(note that what is shown is the logarithm of $\kappa$); however, it is unable to generate a convective zone. In the lower panel of Fig. \ref{058-12-19000-difu} we depict the normalized\footnote{$dW/d\log q$ is normalized to an extremum value of $+1$ or $-1$.} differential work function, $dW/d\log q$ ($q= 1-M_r/M_{\star}$), corresponding to a selected $g$ mode 
with radial order $k= 16$ and period $\Pi= 800$ s. The differential work is negative at the region of the HeII/HeIII opacity bump, therefore this zone contributes to the damping of the mode. 
In this way, this mode turns out to be pulsationally stable. The same goes for all the $g$ modes analyzed for this model. In Fig. \ref{per-teff-eft-059-12} we depict 
the periods of unstable $\ell= 1$ modes  versus $T_{\rm eff}$ corresponding to the same sequence of WD models
($M_{\star}= 0.584 M_{\sun}$, $\log(M_{\rm H}/M_{\star})= -11.8$). Also shown is the $e$-folding time  (in  yr)  of  each  unstable mode, which is defined as $\tau_{\rm e}= 1/|\Im(\sigma)|$, $\Im(\sigma)$,  being $\sigma$ the imaginary part of the complex eigenfrequency. It is an estimate of the time it would take a given mode to reach amplitudes large enough as to be observable. 
Notably, our 
results indicate that all $g$ modes are pulsationally stable at high effective temperatures, so they do not appear in the plot. As the model cools, $g$ modes become unstable due to the $\kappa$ mechanism acting at the partial ionization zone of H 
for effective temperatures lower than $\sim 11\,500$ K, at the instability domain of the ZZ Ceti stars. We conclude that, when the H/He interface is shaped by  element diffusion, there is no excitation of $g$ modes at high effective temperatures due to the partial ionization of  He.

\begin{figure}
	\centering
	\includegraphics[width=1.0\columnwidth]{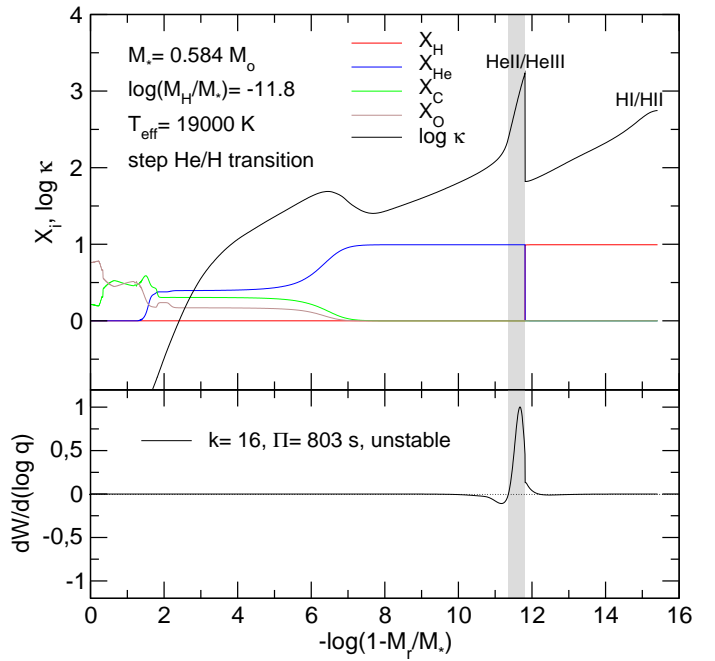}
	\caption{ Upper panel: same as in Fig. \ref{058-12-19000-difu}, but for the extreme case in which the
	H/He interface is assumed to be discontinuous (step-like shape). The vertical gray strip marks the convection zone driven by the $2^{\rm nd}$ ionization of He. Lower panel: the differential work function for an unstable $g$ mode with $k= 16$ and period $\Pi= 803$ s.} 
	\label{058-12-19000-step}
\end{figure}

\begin{figure}
	\centering
	\includegraphics[width=1.0\columnwidth]{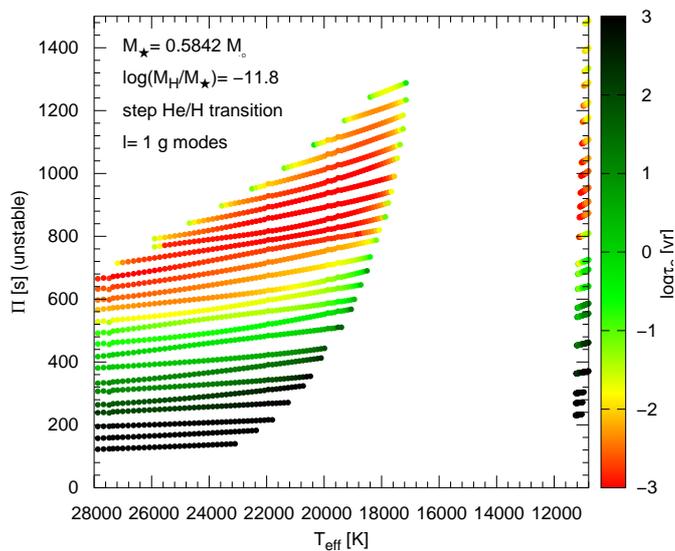}
	\caption{Same as  Fig. \ref{per-teff-eft-059-12}, but for the case in which the
	H/He interface has a step-like shape (see Fig. \ref{058-12-19000-step}). Here, 
	$g$ modes with a wide range of periods are excited at high effective temperatures.}
	\label{per-teff-eft-059-12-step}
\end{figure}

In the upper panel of Fig. \ref{058-12-19000-step}, we display the chemical profiles and the Rosseland opacity for the same WD model, but in this case the H/He transition region is not modeled by element diffusion, but it is assumed to have a step-like 
shape. Here, opacity exhibits a sharp peak at the region of the  H/He chemical interface, thus resulting  in the formation of a narrow convection zone (marked as a gray vertical strip in the figure)  immediately below the H/He chemical interface.
The  peak in $\kappa$ is due to the absence of H below the chemical transition of H/He. We point out that the presence of H in this chemical 
transition when the interface is wide (Fig. \ref{058-12-19000-difu}),   has the effect of reducing the total opacity and suppress convection. This is because, at the He/H transition region, H is fully ionized and its opacity is much lower than that produced by He, which is partially ionized there \citep[see, e.g.,][]{1990ApJS...72..335T}. Hence, when H is absent, the opacity locally reaches much higher values (Fig. \ref{058-12-19000-step}).  In the lower panel of Fig. \ref{058-12-19000-step} we depict the normalized differential work function for the $k= 16$ $g$ mode shown in
the lower panel of Fig. \ref{058-12-19000-difu}. In this case, the differential work has negative values below 
the HeII/HeIII opacity peak, contributing to the damping of the mode, and large positive values at 
the opacity peak, strongly promoting  the driving of the mode. Since driving overcomes damping, the 
mode results globally unstable. The excitation of the mode is due to the $\kappa-\gamma$ mechanism, 
in which the existence of large gradients in the opacity is essential to operate \citep[see, e.g.,][]{2010aste.book.....A}. This is clearly seen in the opacity profile shown in the upper panel of the Fig. \ref{058-12-19000-step}. Fig. \ref{per-teff-eft-059-12-step}  shows 
the period range of excited dipole modes in terms of the effective temperature
for models with step-like shape H/He interface.  At variance with the case depicted in Fig. \ref{per-teff-eft-059-12}, here we find $g$ modes with a
wide range of periods excited in a broad interval of high effective temperatures 
($50\,000\ {\rm K} \gtrsim T_{\rm eff} \gtrsim 17\,000$ K), apart from $g$ modes 
excited at low temperatures in the ZZ Ceti domain. The $e$-folding time has to be compared with the evolutionary timescale ($\tau_{\rm evol}$), that represents the time that the WD spends evolving in the regime of interest.  For our models, $ \tau_{\rm evol}\sim 1.5 \times 10^8$ yr. This is the time it takes for the model to cool from $T_{\rm eff} \sim  23\,000$ K  to $T_{\rm eff} \sim 15\,000$ K. It is much longer than the e-folding times of the excited pulsation modes, $\tau_{\rm e} \lesssim 10^3$ yr. This ensures that the modes have enough time to get excited while the WD models are in the temperature range of interest. For most of the excited modes 
(specifically, for $\Pi \gtrsim 200$ s) the inequality 
$\tau_{\rm e} \ll \tau_{\rm evol}$ is satisfied, which ensures that the modes could reach observable amplitudes as the models are slowly evolving in that range of 
effective temperatures. 

Our results confirm the early findings of 
\cite{1982ApJ...252L..65W} that in DA WD models with very thin H envelopes, $\log(M_{\rm H}/M_{\star}) \leq -10$, He is able to drive $g$-mode pulsations at effective temperatures $T_{\rm eff} \sim 19\,000$ K. Two other relevant works carried out almost at the same time as the analysis by \cite{1982ApJ...252L..65W}, 
that is, the studies by \cite{1981A&A...102..375D} and \cite{1981A&A....97...16D}, reported pulsational instability due to the $\kappa$ mechanism acting at the partial ionization region of the He in models with very thin H envelopes. Nevertheless, the pulsational instabilities they found appear at lower effective temperatures ($T_{\rm eff} \lesssim 13500$ K), compatible with the instability strip of 
ZZ Ceti stars.



\begin{figure}
	\centering
	\includegraphics[width=1.0\columnwidth]{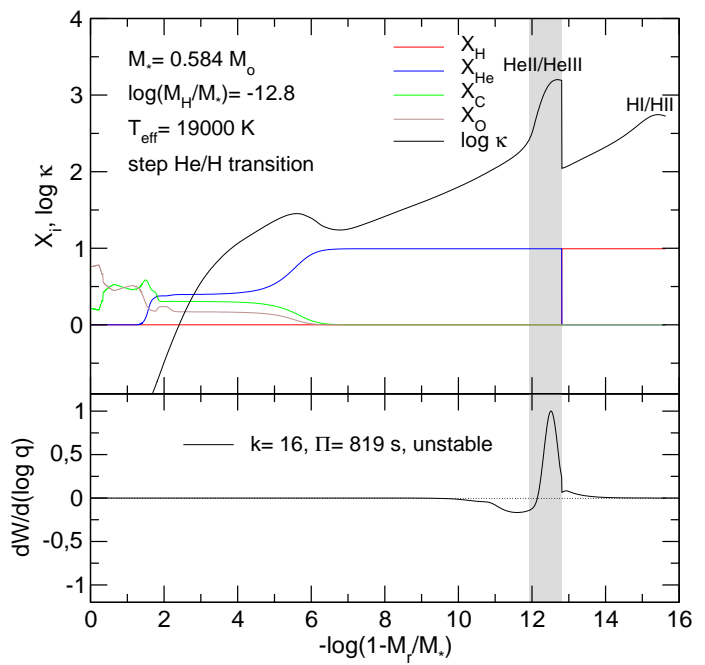}
	\caption{Upper panel: same as in Fig. \ref{058-12-19000-step}, but for the case in which $\log(M_{\rm H}/M_{\star})= -12.8$. Lower panel: the normalized differential work function corresponding to an unstable $g$ mode with $k= 16$ and period $\Pi= 819$ s.} 
	\label{058-13-19000-step}
\end{figure}

\begin{figure}
	\centering
	\includegraphics[width=1.0\columnwidth]{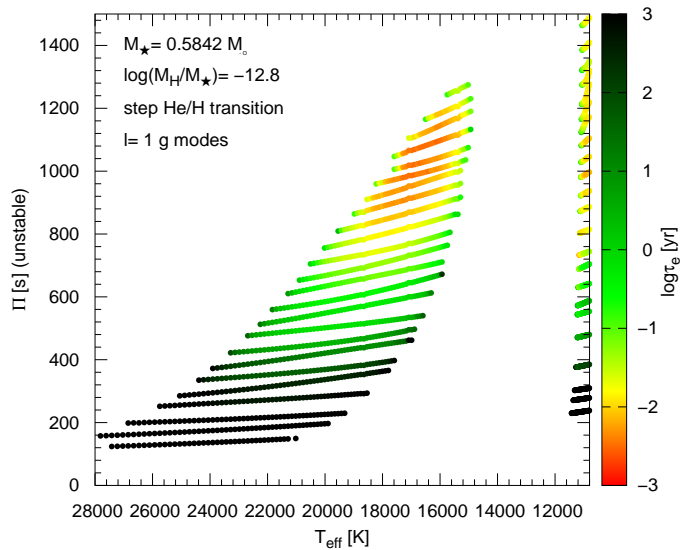}
	\caption{Same as  Fig. \ref{per-teff-eft-059-12-step} but for  $\log(M_{\rm H}/M_{\star})= -12.8$. Again, in this case,
	a lot of $g$ modes are excited at high effective temperatures, apart from the excited 
	modes at the ZZ Ceti phase.} 
	\label{per-teff-eft-059-13-step}
\end{figure}

\begin{figure}
	\centering
	\includegraphics[width=1.0\columnwidth]{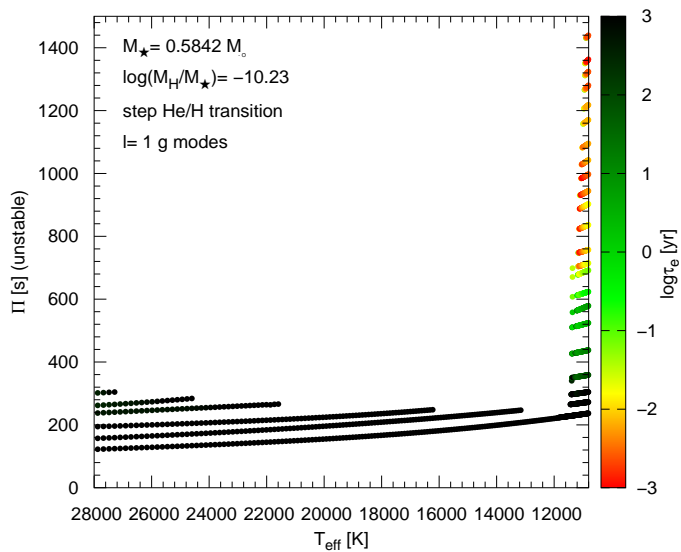}
	\caption{Same as Fig. \ref{per-teff-eft-059-12-step} but for  $\log(M_{\rm H}/M_{\star})=  -10.2$. Here, 	a few low-order modes ($k= 1, \hdots, 6$) are \emph{marginally} unstable apart from the unstable $g$ modes at the ZZ Ceti's 
	effective temperatures.} 
	\label{per-teff-eft-059-10-step}
\end{figure}

Fig. \ref{058-13-19000-step} is similar to Fig. \ref{058-12-19000-step}, but for the case in which $\log(M_{\rm H}/M_
{\star})= -12.8$ and a step-like H/He interface. The unstable $\ell= 1$ $g$-mode periods in terms of $T_{\rm eff}$ for this case are illustrated in Fig. \ref{per-teff-eft-059-13-step}. The results of this Figure are qualitatively similar to the case displayed in Fig. \ref{per-teff-eft-059-12-step} for $\log(M_{\rm H}/M_{\star})= -11.8$, that is, $g$ modes are driven by the $\kappa$ mechanism due to the partial ionization of He for effective temperatures higher than those characterizing ZZ Ceti
stars. We have also analyzed the case in which these models 
harbor a smooth H/He interface modeled by element diffusion. Again, in this case, we do not find unstable modes at high effective temperatures. We also explored the predictions of an even thinner H envelope: $\log(M_{\rm H}/M_{\star})=  -14.5$ with an abrupt H/He transition region. This is the thinnest H envelope considered in this work. Our results for this sequence indicate pulsational instability, but only for $T_{\rm eff} \lesssim 15\,000$ K. Hence, we consider the thickness of this envelope as the lower limit for instability 
pulsational to occur in warm DA WDs.

Fig. \ref{per-teff-eft-059-10-step} displays the 
case of a sequence of models with  the same stellar mass, $M_{\star}= 0.584 M_{\sun}$, but with a thick H envelope of $\log(M_{\rm H}/M_{\star})= -10.2$ having a 
step-like shaped H/He transition region. We find only marginal mode driving of  low order $g$ modes ($k \leq 6$) at high effective temperatures. Indeed, the $e$-folding time of these modes is extremely large, as emphasized by the black color for the period of these modes in Fig.\ref{per-teff-eft-059-10-step}; hence they have no chance to reach observable amplitudes while the  star
is evolving at that stage ($\tau_{\rm e} > \tau_{\rm evol}$). 
On the other hand, we again obtain the unstable $g$ modes 
at the ZZ Ceti stage driven by the partial ionization zone of H via the $\kappa$ mechanism. In practical terms, we can consider that there is no excitation of $g$ modes at high effective temperatures in DA WD models with this H envelope thickness. We have repeated the stability analysis for models with the same thickness of the H envelope but with the H/He transition modeled by diffusion. Again, we do not find excitation of $g$ modes at high effective temperatures.  We conclude that this value of $M_{\rm H}/M_{\star}$ constitutes an upper limit for the H envelope thickness, in such a way that for thicker H envelopes there is no mode driving at high effective temperatures, even in the case in which the H/He transition is characterized by a quasi discontinuous form (step-like shape).

\begin{figure}
	\centering
	\includegraphics[width=1.0\columnwidth]{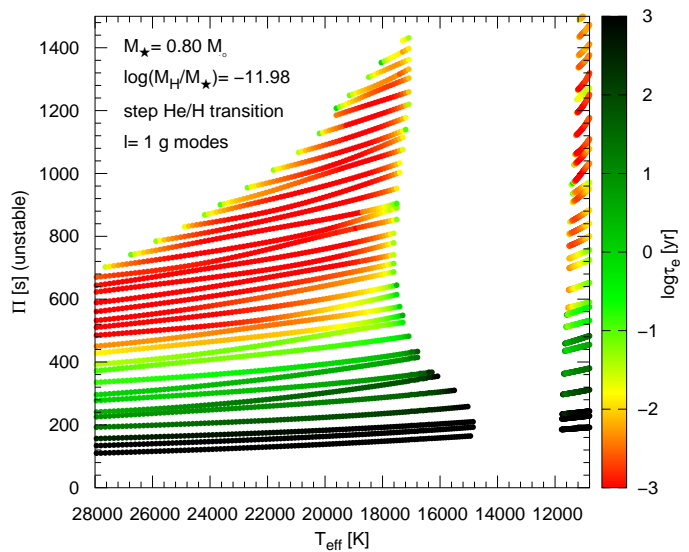}
	\caption{Same as Fig. \ref{per-teff-eft-059-12-step}, but for 
	$M_{\star}= 0.80 M_{\sun}$. Also for these high-mass WD models, many $g$ modes 
	are excited at high effective temperature.} 
	\label{per-teff-eft-080-12-step}
\end{figure}

Finally, we have examined the case of a more massive DA WD model sequence, characterized by $M_{\star}= 0.80 M_{\sun}$ and $\log(M_{\rm H}/M_{\star})= -12$. 
The results of our stability analysis for the  case of an
abrupt H/He transition are illustrated in Fig. \ref{per-teff-eft-080-12-step}. Visibly, for this stellar mass there is also excitation in many $g$ modes for models at high effective temperatures. We also find the instability of $g$ modes in the phase of the ZZ Ceti stars. As in the cases analyzed above, we do not find excitation of modes at high effective temperatures when the H/He chemical transition is the result of element  diffusion (not shown).

\section{Summary and conclusions}
\label{conclusions}

The existence of a new class of pulsating DA WDs hotter than ZZ Ceti stars (warm DAV WDs) was suggested long time ago by  \cite{1982ApJ...252L..65W}, who reported the occurrence of  $g$-mode pulsational instability due to $\kappa$ mechanism acting in the partial ionization zone of He below the H envelope in models of DA WDs with very thin H envelopes ($M_{\rm H}/M_{\star} \lesssim 10^{-10}$), for effective temperatures of $\sim 19\,000$~K.  However, to date, no such warm DAV WDs have been found, despite the large number of pulsating WDs of different types discovered in the last decade both from ground based and space observational surveys. The discovery of warm DAV WDs would constitute a clear demonstration of the existence of DA WDs with  very low H content, thus  shedding light on the evolutionary processes that lead to WD formation in general, particularly the occurrence of late thermal pulses, the most viable scenario to form  WDs with H content lower than $10^{-7} M_\sun$.  The existence of such WDs  cannot be discarded despite the fact that no warm DAV WD has been observed; on the contrary, different pieces of theoretical and observational evidence suggest that a fraction of WDs should be formed with a range of very low H content. 

Our paper has been conducted with the aim of exploring
the impact of  element diffusion   on the stability pulsational properties of warm DAV WDs. We concentrated on full WD evolutionary sequences  of masses 0.80 and 0.58  $M_\sun$ with H content in the range  $M_{\rm H} \sim 10^{-10} -10^{-14}M_\star$, appropriate for the study of pulsational instability of warm DAV WDs. To this end, we have used a new full-implicit treatment for time-dependent element diffusion, that maintains the sum of the chemical abundances
almost constant over large time integrations. This treatment allows us to model the evolution of the outer layer chemical distribution in WDs emerging from VLTP episodes with very low H content in a realistic way, and constitutes a substantial improvement over the semi-implicit treatment considered previously in {\tt LPCODE}.
Initial models were extracted from  progenitor star models that were evolved from the ZAMS, to the thermally-pulsing AGB phase, and through the VLTP occurring at the beginning of the cooling branch \citep{2006A&A...454..845M}.  We find that as WD evolution proceeds, traces of H surviving the VLTP float to the surface, eventually forming thin (and increasing) pure H envelopes and rather extended H/He transition zones.

We have conducted a search for pulsating warm DA WDs among the stars observed by TESS. We have found that 36 DA WD stars with effective temperatures in the range [$17\,500 {\rm K} - 22\,500$ K] have been observed with TESS in s 1 to 9 and we have analyzed them to search for any signs of pulsation. After examining their amplitude spectra,
we conclude that none of the 36 warm DA WD stars observed by TESS show any significant frequency above the $5\sigma$ threshold in the frequency domain of interest. Hence, our search gives null results for pulsating warm DA WDs currently observed by TESS up to the Nyquist frequency at ~4200 $\mu$Hz.

Non-adiabatic pulsations of our warm DAV WD models were computed in the effective temperature range of $30\,000-10\,000$ K. The driving mechanism that can excite pulsations in DA WDs in this regime is the $\kappa$ mechanism due to opacity bumps resulting either from He ionization zone below the H envelope, or 
the H ionization zone, or both  \citep[see the pioneer works by][]{1981A&A...102..375D,1981A&A....97...16D,1982ApJ...252L..65W}. The stability analysis carried out in this work 
was focused on $\ell= 1$ $g$ modes with periods in the range $50-1500$ s. Two possibilities were considered regarding the shape of the H/He chemical interface. On the one hand, we considered this  chemical  transition  as resulting from element diffusion. In this case,
the transition is smooth and wide. On the other hand, we assumed that this transition is extremely narrow and abrupt. Our nonadiabatic results indicate that, in order to have pulsational instability due to the $\kappa$ mechanism acting at the partial ionization zone of He,  it is necessary that the H/He chemical transition region be very narrow and abrupt (see Figs. \ref{058-12-19000-step} and 
\ref{per-teff-eft-059-12-step}). Our computations indicate also that the thickest H envelope for which this kind of mode excitation happens is  $\log(M_{\rm H}/M_{\star})\sim -10$, and the thinnest one is   
$\log(M_{\rm H}/M_{\star})\sim -14.5$. 
On the other hand, if the chemical transition is smooth and extended as element diffusion predicts (see Figs. \ref{058-12-19000-difu} and 
\ref{per-teff-eft-059-12}), then there is no excitation of modes at higher effective temperatures than those characteristic of ZZ Ceti stars, irrespective of the thickness of the H envelope.

We conclude that, assuming that DA WDs with very low H content exist, 
the non-detection of warm DAVs could be attributed to the fact that element diffusion produces a smooth and extended H/He interface, which inhibits the excitation of $g$ modes due to the partial ionization of the He at effective temperatures larger than those typical of ZZ Ceti stars. In this sense, our findings could be considered as an indirect proof that element diffusion indeed operates in the interiors of WDs.
However, it cannot be discarded that the lack of detected pulsations in the TESS data could be attributed to few or none of the warm DA WDs observed in Sectors 1--9 having very thin H envelopes, or that some pulsate with amplitudes below the significance thresholds.

\begin{acknowledgements}
 We thank the anonymous referee for his/her valuable suggestions that improved the content and presentation of the paper. This paper includes data collected with the TESS mission, obtained from the MAST data archive at the Space Telescope Science Institute (STScI). Funding for the TESS mission is provided by the NASA Explorer Program. STScI is operated by the Association of Universities for Research in Astronomy, Inc., under NASA contract NAS 5–26555.
Part of  this work was  supported by  AGENCIA through the  Programa de
Modernizaci\'on    Tecnol\'ogica   BID    1728/OC-AR,   by    the   PIP
112-200801-00940 grant from CONICET,  by MINECO grant AYA2014-59084-P,
by grant G149 from University of La Plata, and by  the AGAUR.  MU acknowledges financial support from CONICYT Doctorado Nacional No. 21190886. ASB gratefully acknowledges financial support from the Polish National Science Center under project No.UMO-2017/26/E/ST9/00703. KJB is supported by an NSF Astronomy and Astrophysics Postdoctoral Fellowship under award AST-1903828. This  research has  made use of  NASA Astrophysics Data System.
\end{acknowledgements}

\bibliographystyle{aa}
\bibliography{hotdav}



\appendix

\section{A sample of warm DA WDs from TESS}
\label{appendixA}

\begin{figure*}
	\centering
	\includegraphics[width=17cm,height=11cm]{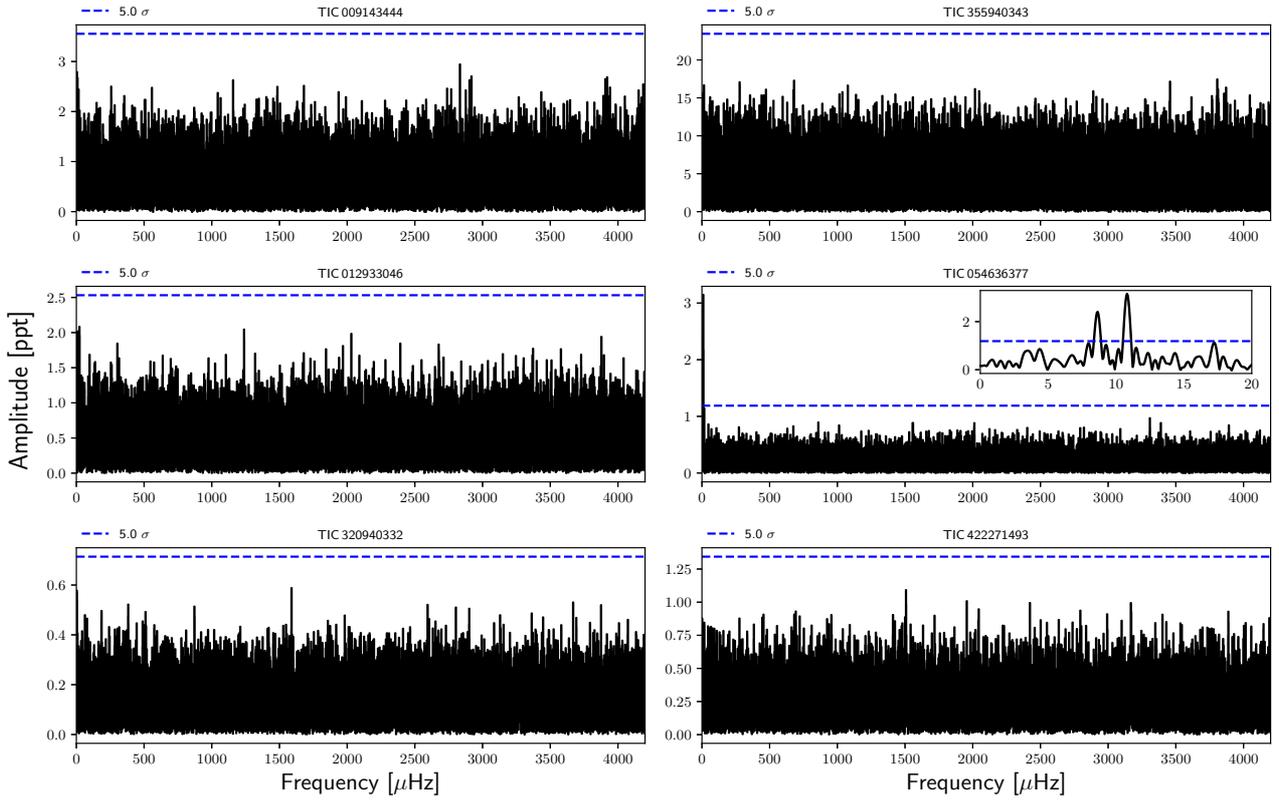}
	\caption{The amplitude spectra of warm DA WDs. The blue dashed horizontal line indicates the threshold (5 $\sigma$). The inset is a blow-up of the amplitude spectra at the low-frequency regime for the few stars where some frequencies are above 5 $\sigma$, as discussed in the text. } 
	\label{ft_1}
\end{figure*}

\begin{figure*}
	\centering
	\includegraphics[width=17cm,height=11cm]{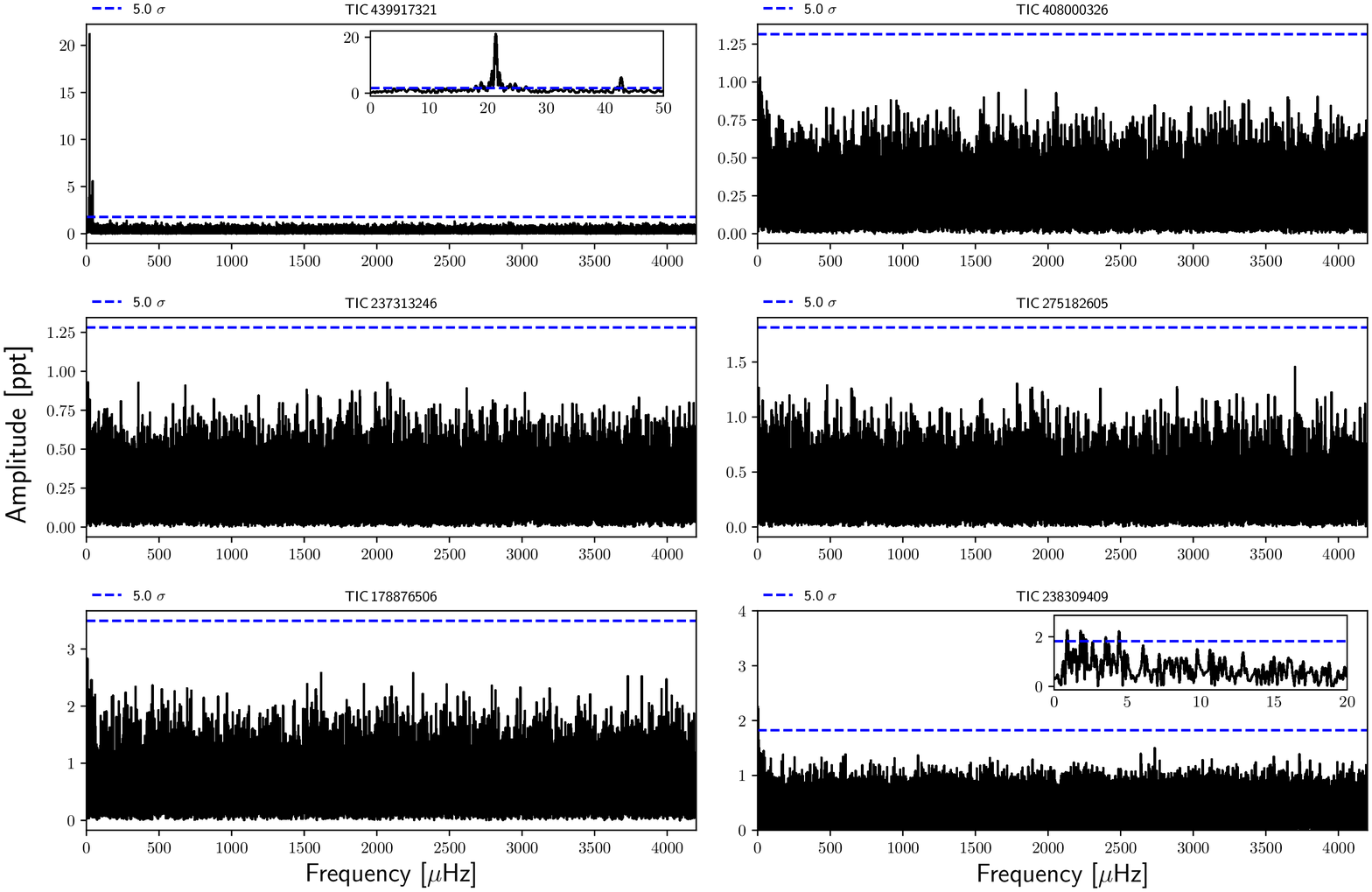}
	\caption{The amplitude spectra of warm DA WDs (cont.). }
	\label{ft_2}
\end{figure*}

\begin{figure*}
	\centering
	\includegraphics[width=17cm,height=11cm]{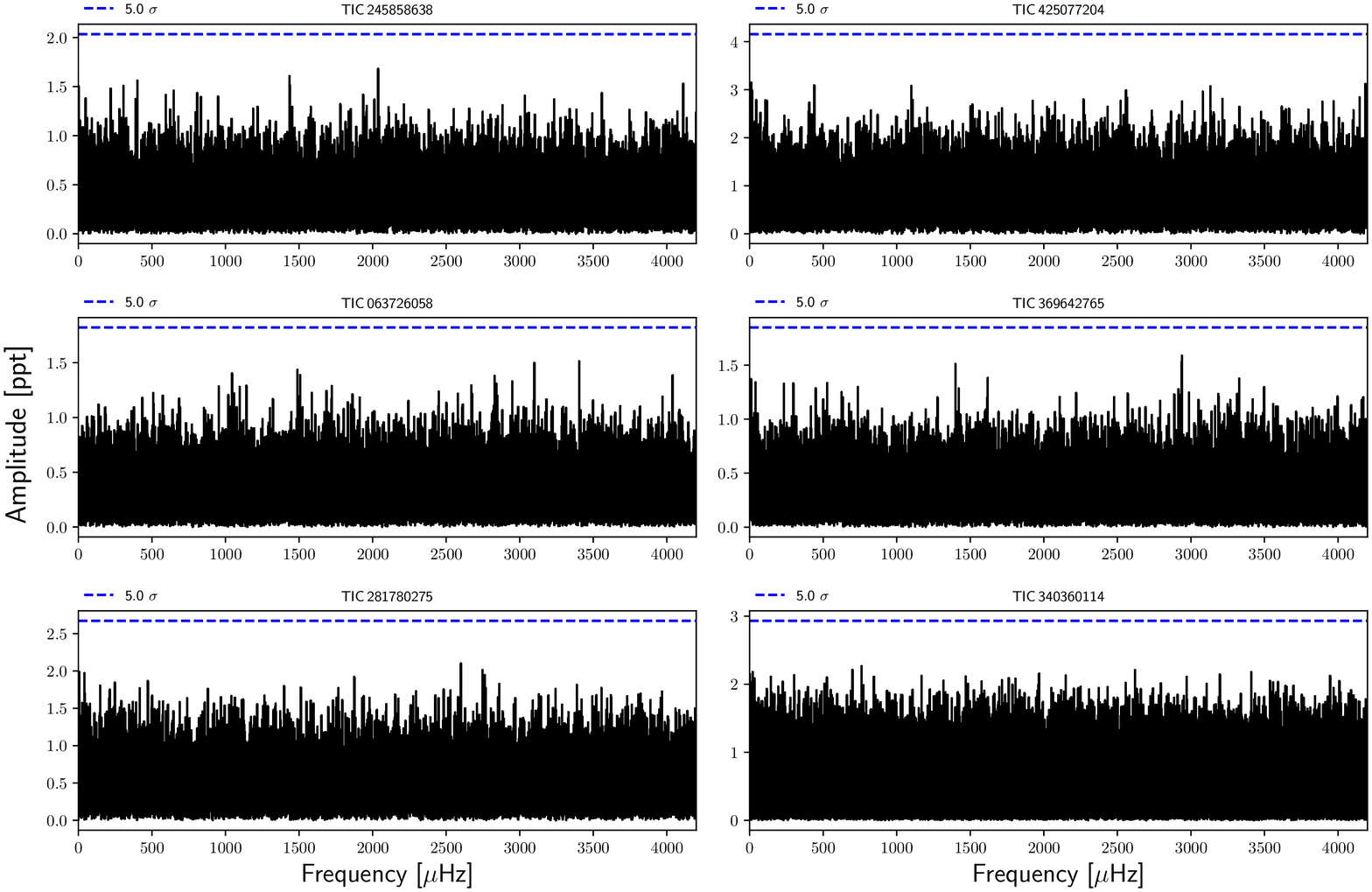}
	\caption{The amplitude spectra of warm DA WDs (cont.). }
	\label{ft_3}
\end{figure*}

\begin{figure*}
	\centering
	\includegraphics[width=17cm,height=11cm]{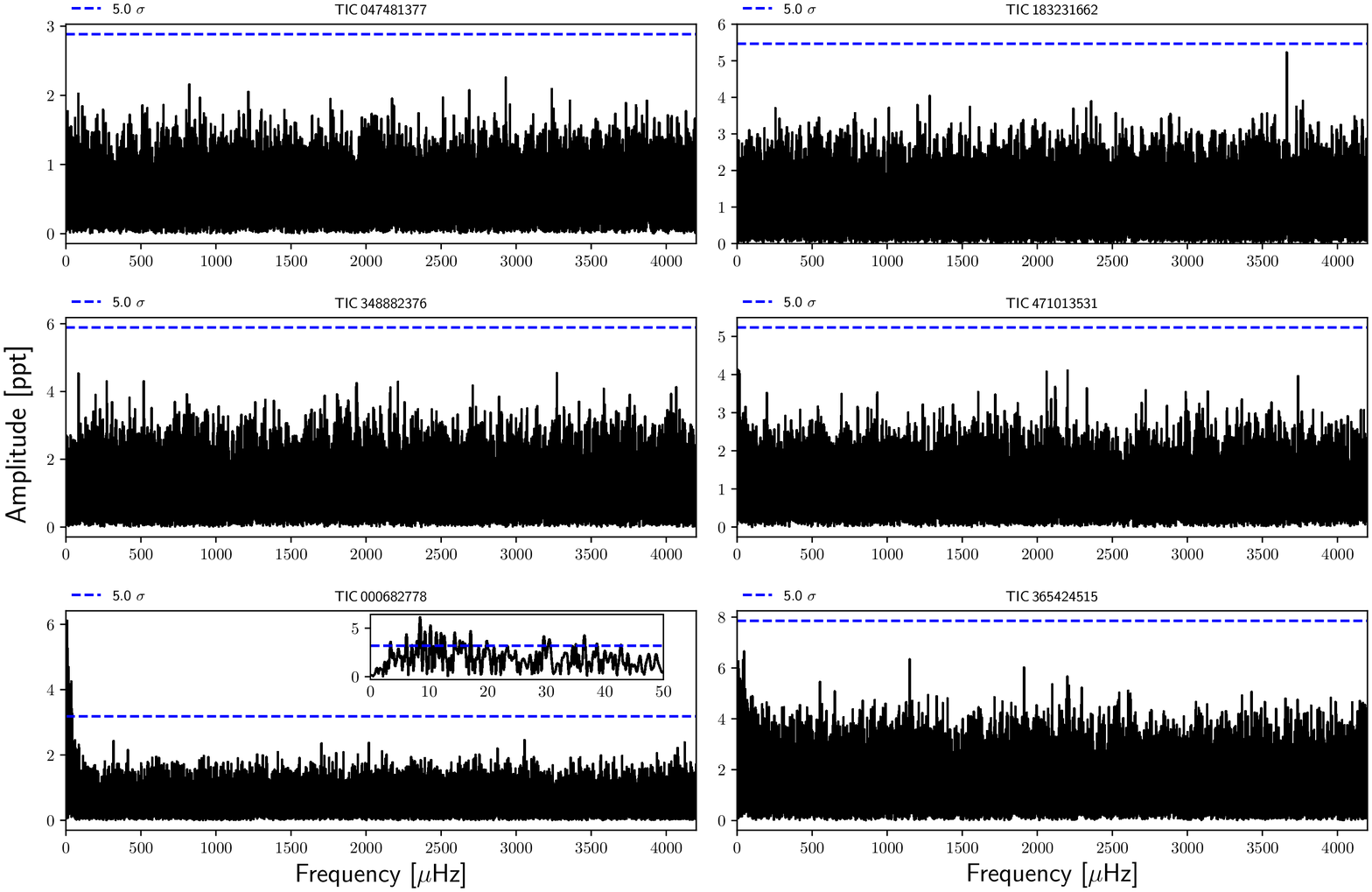}
	\caption{The amplitude spectra of warm DA WDs (cont.).}
	\label{ft_4}
\end{figure*}

\begin{figure*}
	\centering
	\includegraphics[width=17cm,height=11cm]{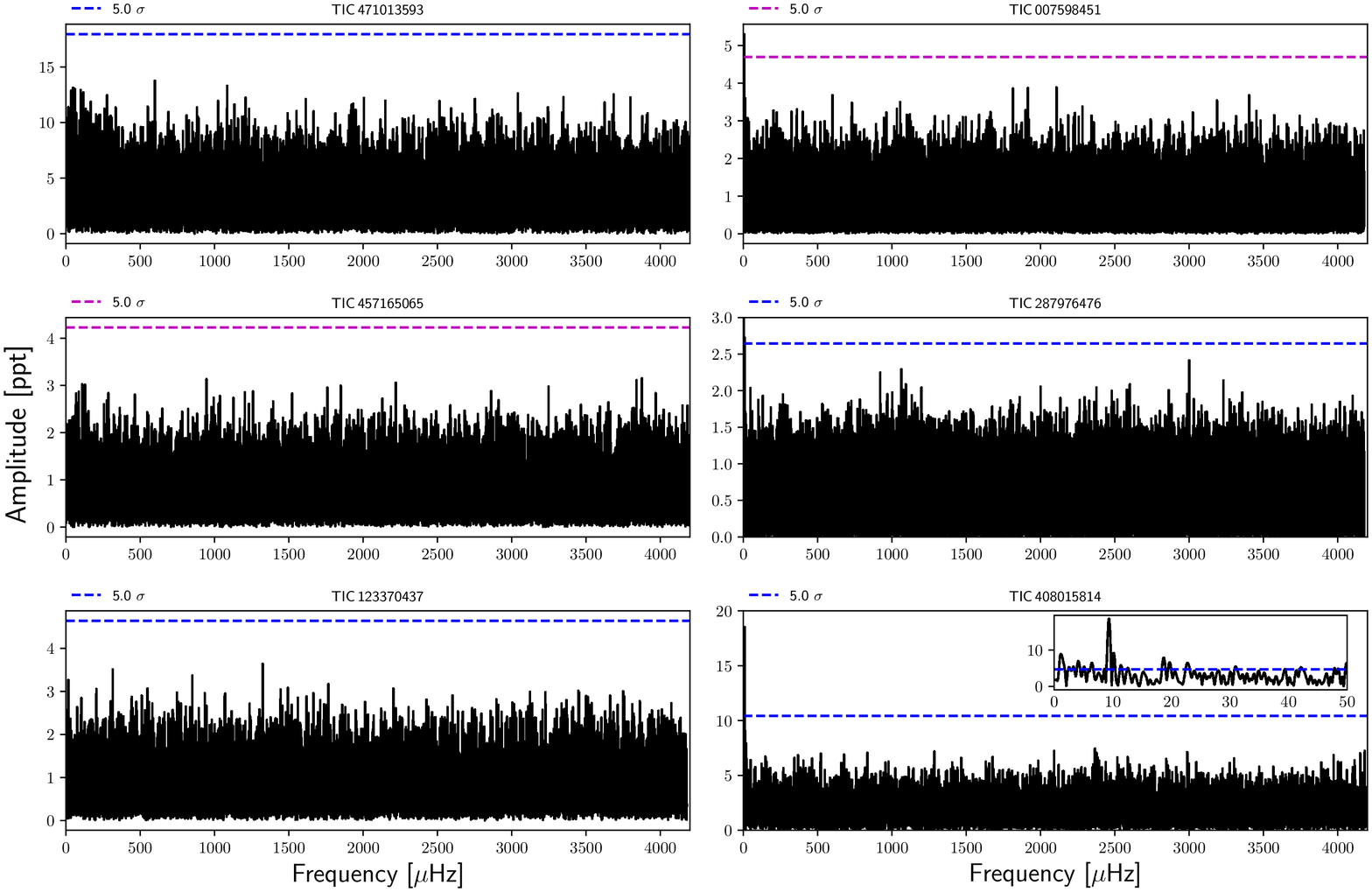}
	\caption{The amplitude spectra of warm DA WDs (cont.). }
	\label{ft_5}
\end{figure*}

\begin{figure*}
	\centering
	\includegraphics[width=17cm,height=11cm]{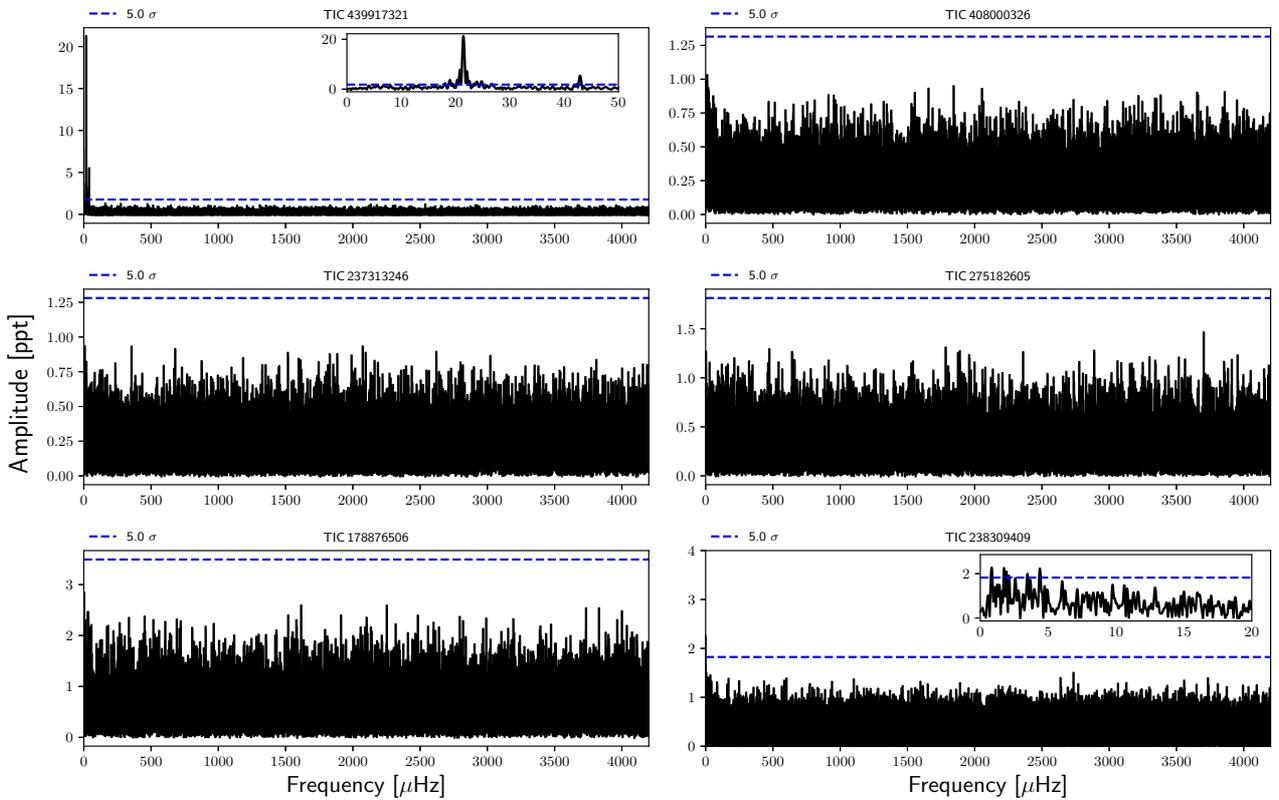}
	\caption{The amplitude spectra of warm DA WDs (cont.).}
	\label{ft_6}
\end{figure*}

Here, we show  
the Fourier Transforms of the sample of 36 warm DA WDs observed with TESS mission. Table \ref{table1} lists the complete sample of stars analysed in this paper including the spectral types, effective temperatures and surface gravities (taken from the MWDD Catalog), along with the TESS magnitudes, observed sector, average noise level and the adopted threshold ($5\sigma$). In Figs. 
\ref{ft_1} to \ref{ft_6} we depict the Fourier Transforms of the stars listed in table \ref{table1}. As can be seen, none of the analyzed stars exhibit signals of intrinsic variability in the frequency range of expected pulsation. The peaks seen in few stars at very low frequencies that are above $5\sigma$ threshold (TIC 408015814, TIC 054636377, TIC 237313550, TIC 000682778) are low-frequency systematic noise that has been reported in some of the TESS data. 
The only exception is the star TIC\,439917321. This is a member of WD$+$dM (M dwarf) binary system, with a period of 12.98 hours. In the FT of TIC\,439917321, the peaks  $f_1$ and $f_2$ (see Table \ref{table-TIC439917321}) are signatures of binarity. After pre-whitening the orbital frequency and it's harmonic ($f_1$ and $f_2$), there is no other periodic signal detected above the 
threshold ($5\sigma=1.9432$) and we conclude that TIC\,439917321 does not show any signs of pulsation.

\begin{table*}
\caption{The list of complete set of 36 warm DA WD targets observed with TESS (sectors 1 to 9) including 3 different sets of information along with their names (columns 1 and 2). The first set consists of the fundamental parameters of targets from the MWDD Catalog (columns 3, 4, and 5). The second set of parameters comes from TESS mission including magnitudes and observed sector which is analysed in this paper (columns 6 and 7). Finally, the third set consists of average noise level and the threshold ($5\sigma$) (columns 8 and 9). 
}
\begin{tabular}{cccccccccc}
\hline
TIC & Name & S. Type & $T_{\rm eff}$ & $\log g$ & $T_{\rm mag}$ & Obs. Sector & Average Noise Level & Threshold \\
    &      &        & [K]           & [cm/s$^2$] &     &    &  [ppt] & [ppt]  \\
\hline
009143444 & WD2322$-$181 & DA & 21862 & 8.009 & 10.29 & 2 & 0.759 & 3.554 \\
355940343 & WD0850$-$617 & DA & 21840 & 7.99 & 13.41 & 9 $-$ 10 & 4.987 & 23.449  \\
012933046 & WD 2259$-$267 & DA & 20050 & 8.03 & 13.71 & 2 & 0.543 & 2.531  \\
054636377 & EGGR 57 & DA & 19440 & 7.92 & 13.79 & 7 & 0.254 & 1.188 \\
320940332 & LP 426$-$26 & DA & 19211 & 7.959 & 13.85 & 1 & 0.152 & 0.714 \\
422271493 & WD 1049$-$158 & DA & 18085 & 8.321 & 13.87 & 9 & 0.286 & 1.343  \\
439917321 & PSOJ030607 & DA+M & 19193 & 7.9 & 14.13 & 4 & 0.389 & 1.773  \\
408000326 & EGGR 31 & DA & 19637 & 7.933 & 14.25 & 5 & 0.281 & 1.314 \\
237313246 & WD 2251$-$634 & DA & 20340 & 7.95 & 14.27 & 1 & 0.272 & 1.281  \\
275182605 & PG 0933+026 & DA & 20987 & 7.821 & 14.27 & 8 & 0.386 & 1.813 \\
178876506 & PG 1003$-$023 & DA & 18040 & 7.842 & 14.55 & 8 & 0.744 & 3.493 \\
238309409 & BPM18764 & DA & 21970 & 7.82 & 14.71 & 7$-$8$-$9 & 0.390 & 1.823 \\
245858638 & EGGR 39 & DA & 20799 & 8.104 & 14.73 & 5 & 0.432 & 2.034 \\
425077204 & WD 0242$-$174 & DA & 20215 & 7.954 & 14.99 & 4 & 0.883 & 4.155 \\
063726058 & GD 603 & DA & 22160 & 8.508 & 15.01 & 2 & 0.387 & 1.821 \\ 
369642765 & WD 0047$-$524 & DA & 18831 & 7.89 & 15.01 & 2 & 0.393 & 1.848\\
281780275 & WD0048$-$544 & DA & 17870 & 7.976 & 15.21 & 2 & 0.570 & 2.671  \\
340360114 & WD 0740$-$57 & DA & 21670 & 8.17 & 15.22 & 7$-$8$-$9$-$10 & 0.624 & 2.931 \\
047481377 & WD1020$-$207 & DA & 18491 & 7.86 & 15.29 & 9 &  0.611 & 2.884  \\
183231662 & MCT2345$-$3940 & DA & 19197 & 7.866 & 15.43 & 2 & 1.166 & 5.465  \\
348882376 & WD 0951$-$155 & DA & 18590 & 8.01 & 15.52 & 8 & 1.254 & 5.890  \\
471013531 & GD1352 & DA & 19901 & 7.677 & 15.58 & 3 & 1.112 & 5.234  \\
000682778 & WD 0453$-$295 & DA & 20640 & 7.61 & 15.71 & 5 & 0.689 & 3.181 \\
365424515 & HS 0309+1001 & DA2.7 & 18786 & 7.725 & 16.16 & 4 & 1.671 & 7.853 \\
471013593 & FBS 0341$-$008 & DA & 22588 & 7.617 & 16.18 & 5 & 3.749 & 17.569 \\
007598451 & HE0414$-$4039 & DA2.4 & 21664 & 8.091 & 16.39 & 4$-$5 & 0.945 & 4.438  \\
457165065 & WD 0341+021 & DA & 18117 & 7.132 & 16.71 & 5 & 0.874 & 4.108  \\ 
287976476 & BPM5102 & DA & 18180 & 7.87 & 16.79 & 6$-$10$-$11$-$12 & 0.563 & 2.644  \\
123370437 & GD1442 & DA & 17810 & 7.95 & 17.56 & 3 & 0.980 & 4.593  \\
408015814 & WD 0053$-$090 & DA & 22372 & 6.147 & 17.62 & 3 & 2.217 & 10.424 \\
296860904 & WD 1031$-$147 & DA & 22119 & 7.597 & 17.83 & 9 & 1.593 & 7.498  \\
237313550 & HE2251$-$6218 & DA & 18033 & 7.827 & 18.07 & 1 & 0.940 & 4.395  \\
419013508 & GD765 & DA & 18335 & 7.821 & 18.19 & 3 & 1.255 & 5.884  \\
001354950 & HE0452$-$3444 & DA2.5 & 21810 & 7.887 & 18.72 & 5 & 1.018 & 4.790  \\
101100431 & HE0348$-$4445 & DA & 19951 & 8.069 & 18.83 & 3$-$4 & 1.536 & 7.212  \\
178835540 & LB 564 & DA & 18722 & 7.893 & 19.33 & 8 & 1.676 & 7.875 \\

\hline
\label{table1}
\end{tabular}
\end{table*}

\begin{table}
\caption{Frequencies detected in TIC\,439917321.}
\begin{tabular}{lcc}
\hline
\noalign{\smallskip}
Peak & Frequency    & Amplitude \\
     &   [$\mu$Hz]  &   [ppt]   \\ 
\noalign{\smallskip}
\hline
\noalign{\smallskip}
$f_1$ &   $21.3915    \pm 0.0037$  & $21.13 \pm 0.34$ \\
$f_2$ &   $42.7931    \pm 0.0152$  & $5.12 \pm 0.34$ \\ 
\noalign{\smallskip}
\hline
\end{tabular}
\label{table-TIC439917321}
\end{table}










\section{Numerical treatment of time-dependent element diffusion}
\label{appendixB}

Under the  influence of  gravity, partial pressure, and   induced  electric   fields, the   diffusion  velocities   in  a multicomponent  plasma satisfy  the  set of  $N-1$  independent  
 equations (Burgers 1969):

\begin{equation}
\frac{dp_i}{dr}-\frac{\varrho_i}{\varrho}\frac{dp}{dr}-n_iZ_ieE=
\sum_{j\ne i}^NK_{ij}\left(w_j-w_i\right),
\label{diff1}
\end{equation}

\noindent where, $p_i$, $\varrho_i$, $n_i$, $Z_i$, and $w_i$ denote,
respectively, the partial pressure,  mass density, number density,
mean charge, and diffusion velocity for  chemical species  $i$. $N$ is  the number of  ionic species
plus  electrons. The unknown  variables  are  $w_i$ and the electric field
$E$. Resistance  coefficients $K_{ij}$ are from \cite{1986ApJS...61..177P}. This set of equations is solved together with the equations  for  no  net mass  flow $\sum_iA_in_iw_i=0$ and no electrical current $\sum_iZ_in_iw_i=0$. In the evolutionary sequences computed in this work, thermal diffusion has been
neglected. This does not change the conclusions of the paper. We check it out
by including thermal diffusion in the computation of the  
$M_{\star}= 0.584 M_{\sun}$ sequence with  $\log(M_{\rm H}/M_{\star})= -11.8$. 
Although the inclusion of thermal diffusion leads to a thicker pure H envelope and a steeper H/He transition region, no excitation of $g$ modes at high effective temperatures due to the partial ionization of  He is found, see text.

Following \cite{1985ApJ...296..540I}, Eq.~(\ref{diff1}) can be rearranged to obtain (for the ions)

\begin{equation}
\frac{1}{n_i}\sum_{j\ne i}^NK_{ij}\left(w_i-w_j\right) - Z_ieE=
\alpha_i-k_{\rm B}T\frac{d\ln n_i}{dr},
\end{equation}

\noindent where $\alpha_i= -A_i\ m_{\rm H}\ g - k_{\rm B}\ T\ (d \ln T/dr) $,  being  $g$  the  gravitational acceleration. $w_i$  can be written

\begin{equation}
  w_i=w^{\rm gt}_i-\sum_{j}\sigma_{ij}\frac{d\ln n_j}{dr},
  \label{velocidad}
\end{equation}

\noindent where  $w^{\rm gt}_i$  stands  for  the velocity  component
due to gravity and temperature gradient (gravitational settling component).
The components due to gradients in number density are referred to as chemical diffusion components.  The
summation is  done {\sl  for the ions  only}. This  equation, together
with the conditions for no net mass flow and electrical current,  can be
solved  using matrix
inversion to  find $w_{i}^{\rm gt}$ and  $\sigma_{ij}$, which finally
are employed to follow the evolution of the number density of ionic species $i$


\begin{equation}
\frac{\partial n_i}{\partial t}=-\frac{1}{r^2}\frac{\partial}{\partial r}
\left(r^2\ n_i\  w_{i} \right).
\end{equation}

For numerical purposes, we write this equation in terms of the abundance by mass $X_i$ of species $i$, given by $X_i=\varrho_i / \varrho$

\begin{equation}
\varrho\ \frac{\partial X_i}{\partial t}=-\frac{1}{r^2}\frac{\partial}{\partial r}
\left(r^2\ X_i\ \varrho\  w_{i} \right).
\end{equation}

We multiply this equation by $4\pi r^2$ and integrate over the radial coordinate $r$ from
$r-\Delta r/2$ to $r+\Delta r/2$, to obtain

\begin{equation}
\Delta m {\partial X_i \over \partial t}\bigg|_{r}=-\left( 4\pi \varrho r^2 X_i  w_{i} \right ) \bigg|_{r-\Delta r/2}^{r+\Delta r/2},
\label{fflu} 
\end{equation}

\noindent where $\Delta m=\int_{r-\Delta r/2}^{r+\Delta r/2} 4 \pi r^2 \varrho dr$ (we assume  $\Delta r$ small enough for $\partial X_i / \partial t$ to be constant over  $\Delta r$). Eq. (\ref{fflu}) simply states
 that the difference of flux of mass of
 element $i$ between the two borders of a spherical shell is equal
 to the time variation of the mass of element $i$ in the shell. We discretize this equation with respect to time and space by dividing the star into $K$ mesh points, with the lagrangian mass coordinate of the $k^{th}$ mesh point being $m_k$ (the mass inside the sphere of radius $r_k$). In what follows, all the quantities with an indice $k$ are evaluated at the mesh point $k$. The mesh points are numbered from 1 at the surface to $K$ at the centre. The mass of element $i$ removed from mesh point $k+1$ and added to mesh point $k$ per unit time becomes:

\begin{equation}
4 \pi r^2_{k+1/2} \varrho_{k+1/2} (X_i w_i)_{k+1/2},
\end{equation}

\noindent where the subscript $k+1/2$ means that quantities are evaluated at the
midpoint between the mesh points $k+1$ and $k$. For instance,  
$r_{k+1/2} =(r_{k}+r_{k+1})/2$. We have a similar expression
for the mass of element $i$ removed from mesh point $k$ and added to mesh point $k-1$.
Denoting with  $X_{i,k}$  the abundance by mass of element $i$ evaluated at the mesh point $k$ and  considering that $\Delta m_k=\int_{r_{k+1/2}}^{r_{k-1/2}} 4 \pi r^2 \varrho dr$, we have

\begin{eqnarray}
\lefteqn{\Delta m_k {\partial (X_{i,k}) \over \partial t}= 4 \pi r^2_{k+1/2}\ \varrho_{k+1/2}\ (X_i w_i)_{k+1/2}} \nonumber
\\ & &
 -\ {4 \pi r^2_{k-1/2}\ \varrho_{k-1/2}\ (X_i w_i)_{k-1/2}}.
 \label{flujo}
\end{eqnarray}

The left hand term of this equation  is discretized as

\begin{equation}
 \Delta m_k {X_{i,k}-X_{i,k}^{0} \over \Delta t},
\end{equation}

\noindent where $X_{i,k}^{0}$ corresponds to the abundance by mass of element $i$ at mesh point $k$ evaluated at the beginning of time step $ \Delta t$, and $X_{i,k}$ the abundance of the same element at $k$ at the end of the time step. Note that we are considering
the average  of the product $X_i\ w_i$ (specifically, $X_i\ w_{i}^{\rm gt}$ and  $X_i\ \sigma_{ij}$) between two adjacent mesh points and not the product of the averages of $X_i$ and  $w_i$. In this way,   mass conservation is guaranteed in Eq.\ref{flujo}, as we imposed the condition of no net mass flow at each mesh point $k$  at calculating
the diffusion velocities\footnote{For instance, 
\begin{eqnarray}
 \sum_i (X_i
 w_i)_{k+1/2}=\sum_i \frac{ X_{i,k} w_{i,k}+X_{i,k+1}
  \nonumber
  w_{i,k+1}}{2}\propto \underbrace{\sum_i A_i  n_{i,k} w_{i,k}}_{=0}\\ +
  \underbrace{\sum_i   A_i  n_{i,k+1}   w_{i,k+1}}_{=0}=0
  \nonumber
\end{eqnarray}
}.

Taking $\Delta m_k=\varrho_k\ \Delta V_k $, with $\Delta V_k=4 \pi (r^3_{k-1/2} -r^3_{k+1/2})/3$ we obtain

\begin{eqnarray}
\lefteqn{{X_{i,k}-X_{i,k}^{0} }
  ={\Delta t \over{\varrho_k\ \Delta V_k}}\ 4 \pi r^2_{k+1/2}\ \varrho_{k+1/2}\ (X_i w_i)_{k+1/2}}\nonumber
\\ & &
 -\ {{\Delta t \over{\varrho_k\ \Delta V_k}}\ 4 \pi r^2_{k-1/2}\ \varrho_{k-1/2}\ (X_i w_i)_{k-1/2}},
\label{diff2}
\end{eqnarray}

\noindent where (from Eq.  \ref{velocidad})

\begin{eqnarray}
\lefteqn{{ (X_i w_i)_{k+1/2}}
  =\frac{X_{i,k} w_{i,k}^{\rm gt} +X_{i,k+1} w_{i,k+1}^{\rm gt}}{2}\ - \sum_{j}\frac{\sigma_{ij,k} X_{i,k} + \sigma_{ij,k+1} X_{i,k+1}}{2}}\nonumber
\\ & & 
\frac{2}{X_{j,k}+X_{j,k+1}}\ \frac{\varrho_kX_{j,k}-\varrho_{k+1} X_{j,k+1}}{\varrho_{k+1/2}(r_k-r_{k+1})}
\label{diff3}
\end{eqnarray}

\noindent and

\begin{eqnarray}
\lefteqn{{ (X_i w_i)_{k-1/2}}
  =\frac{X_{i,k} w_{i,k}^{\rm gt} +X_{i,k-1} w_{i,k-1}^{\rm gt}}{2}\ - \sum_{j}\frac{\sigma_{ij,k} X_{i,k} + \sigma_{ij,k-1} X_{i,k-1}}{2}}\nonumber
\\ & & 
\frac{2}{X_{j,k}+X_{j,k-1}}\ \frac{\varrho_{k-1}X_{j,k-1}-\varrho_{k} X_{j,k}}{\varrho_{k-1/2}(r_{k-1}-r_k)}
\label{diff4}
\end{eqnarray}

We have a set of difference equations at each mesh point $k$, the solution of which will give us $X_{i,k}$ for all species $i$ at each $k$. The boundary conditions at the center, $k=K$, and at the surface, $k=1$, are respectively

\begin{eqnarray}
\Delta m_K {\partial (X_{i,K}) \over \partial t}=  - {4 \pi r^2_{K-1/2} \varrho_{K-1/2} (X_i w_i)_{K-1/2}}
\end{eqnarray}

\begin{eqnarray}
\Delta m_1 {\partial (X_{i,1}) \over \partial t}=   {4 \pi r^2_{1+1/2} \varrho_{1+1/2} (X_i w_i)_{1+1/2}},
\label{super}
\end{eqnarray}

\noindent for which we proceed as above. Eqs. (\ref{diff2}) to (\ref{super}) can be written in a compact form as

\begin{equation}
\left\lbrace
\begin{array}{lll}
  B_{i}(\vec{X_{\rm k}},\, \vec{X_{\rm k+1}})=0 & k=1 \\
  G_{i,k}(\vec{X_{\rm k-1}},\, \vec{X_{\rm k}},\, \vec{X_{\rm k+1}})=0 &  2 \leq k \leq K-1 \\
  C_{i}(\vec{X_{\rm k-1}},\, \vec{X_{\rm k}})=0  & k=K
\end{array}
\right.
\label{set} 
\end{equation}

\noindent where $\vec{X_{\rm k}}$ means $X_{l,k}\,  \{l=1, \cdots, N_{\rm ions} \}$ with $N_{\rm ions}$ the number of chemical species. 
Here, $B_i$, $G_{i,k}$, and $C_i$ are given by

\begin{eqnarray}
B_i= \lefteqn{{X_{i,1}-X_{i,1}^{0} } 
- {\Delta t \over{\varrho_1\ \Delta V_1}}\ 4 \pi r^2_{1+1/2}\ \varrho_{1+1/2}\ (X_i w_i)_{1+1/2}}\nonumber
\\ & &
\label{b_i}
\end{eqnarray}

\noindent where $(X_i w_i)_{1+1/2}$ is given by Eq. (\ref{diff3}) with $k= 1$,

\begin{eqnarray}
C_i= \lefteqn{{X_{i,K}-X_{i,K}^{0} } 
+ {\Delta t \over{\varrho_K\ \Delta V_K}}\ 4 \pi r^2_{K-1/2}\ \varrho_{K-1/2}\ (X_K w_i)_{K-1/2}}\nonumber
\\ & &
\label{c_i}
\end{eqnarray}

\noindent where $(X_K w_i)_{K-1/2}$ is given by Eq. (\ref{diff4}) with $k= K$, and   

\begin{eqnarray}
G_{i,k}= \lefteqn{{X_{i,k}-X_{i,k}^{0} } 
- {\Delta t \over{\varrho_k\ \Delta V_k}}\ 4 \pi r^2_{k+1/2}\  \varrho_{k+1/2}\ (X_i w_i)_{k+1/2}}\nonumber 
\\ & &
+\lefteqn{{\Delta t \over{\varrho_k\ \Delta V_k}}\ 4 \pi r^2_{k-1/2}\ \varrho_{k-1/2}\ (X_i w_i)_{k-1/2}}
\label{g_i}
\end{eqnarray}

\noindent where $(X_i w_i)_{k+1/2}$ and $(X_i w_i)_{k-1/2}$ are given by Eq. (\ref{diff3}) and Eq. (\ref{diff4}), respectively.

To solve this set of non-linear equations we use an iterative full implict method, in which an initial solution is gradually improved by applying successive simultaneous corrections to all variables at all mesh points. Implicit methods, in constrast with semi-implicit methods, are usually preferred because of the superior stability characteristics. In addition, they maintain the sum of $X_i$ almost constant over large time integrations. This is a relevant issue, since in this work we are interested in the formation and evolution of  H envelopes  less massive than $10^{-10}M_\sun$. The treatment present here is an improvement over that  presented in 
\cite{2003A&A...404..593A}, based on the semi-implicit method  described in \cite{1985ApJ...296..540I}. 

Let  $\vec{X}^1$ be a first approximation to the set of  Eqs. \ref{set}, which as a
first guess we take it as the solution at the previous time $t-\Delta t$. Clearly, since this is only an approximation to the solution, then $B_{i}(1) \neq 0$,  $G_{i,k}(1) \neq 0$, and $ C_{i}(1) \neq 0$. Let  $\delta \vec{X}$ the corresponding corrections to all variables at all mesh points; we get a second approximation  $ \vec{X}^{2}=  \vec{X}^1 + \delta \vec{X}$, so that now $B_{i}(2) = 0$,  $G_{i,k}(2) = 0$, and $ C_{i}(2) = 0$. If corrections are small, a first-order Taylor expansion leads 

\begin{equation}
 \sum_{j}\frac{\partial G_{i,k}}{\partial X_{j,k-1}} \delta X_{j,k-1} + 
\sum_{j}\frac{\partial G_{i,k}}{\partial X_{j,k}} \delta X_{j,k} +
\sum_{j}\frac{\partial G_{i,k}}{\partial X_{j,k+1}} \delta X_{j,k+1}= -G_{i,k}(1)
\end{equation}

\noindent and similarly for $B_{i}$ and  $ C_{i}$. The derivatives are assessed at  the first
approximation. From here, the corrections $\delta \vec{X}$ are found by inverting a
band-type matrix  with non-vanishing coefficients only in blocks
near the diagonal (because difference equations depend only on variables at
adjacent points), by following the elimination procedure of  \cite{1964ApJ...139..306H}.
After several iterations (diffusion velocities are 
 re-calculated at each iteration), the approximate solution is improved until the
absolute value of all corrections drops below a given limit or the difference equations
are satisfied to a given accuracy.

This implicit procedure for solving the time-dependent diffusion equations is
efficient and stable. In particular, for our sequences with $M_{\rm H}=10^{-11} M_\sun$, the total mass of the H content changes by less than  0.01$\%$ after $10^8$ yr of evolution.  After the convergence of a stellar model, the chemical abundance changes due to element diffusion  and then due to convection (and nuclear reactions if any) for the next evolutionary time step are evaluated, and then the next stellar model is computed by solving the full set of stellar structure and evolution equations.


\end{document}